\documentclass{article}

\usepackage{amssymb,amsfonts,amsmath}
\usepackage{cite,enumerate,float}
\usepackage{color}
\usepackage{tikz}
\usetikzlibrary{arrows,snakes,backgrounds}

\def\be{\begin{eqnarray}}
\def\ee{\end{eqnarray}}

\def\p{\partial}
\def\tr{{\rm tr}\,}
\def\Tr{{\rm Tr}\,}

\definecolor{red}{rgb}{1,0,0}
\definecolor{orange}{rgb}{1,0.5,0}
\definecolor{violet}{rgb}{0.7,0,1}

\hoffset= - 1.0in         

\begin{document}

\hfill MIPT/TH-14/22

\hfill ITEP/TH-17/22

\hfill IITP/TH-15/22

\bigskip

\centerline{\Large{
Integrability and Matrix Models
}}

\bigskip

\bigskip

\centerline{{\bf A.Morozov}}

\bigskip

\centerline{\it NRC "Kurchatov Institute", ITEP, MIPT \& IITP, Moscow, Russia}

\bigskip

\centerline{ABSTRACT}

\bigskip

{\footnotesize
A brief review of the eigenvalue matrix model integrability and superintegrability properties,
focused on the simplest, still representative, Gaussian Hermitian case.

}

\bigskip

\bigskip

\section{Introduction}

The theory of the (eigenvalue) matrix models (MAMO) occupies more and more space in the
context of modern theoretical physics,
what could seem unexpected and even unjustified.
However, this impression is basically wrong:
MAMO are indeed crucially significant, because they appear to capture
the most interesting theoretical properties of generic field/string theory.
Most important, they allow to study the "non-perturbative" properties
of integrals and their interplay with the group/symmetry structures --
avoiding additional (and rarely significant) complexities of {\it functional}
integration.
The main property of exact non-perturbative partition function is {\it integrability},
which reflects invariance of the (functional) integral under arbitrary change
of integration variables (fields).
It appears related to the set of equations for partition function
as a function of (an infinite set of) the coupling constants,
in particular, to the definition of an integral with parameters (couplings)
as a $D$-module.
$D$-module is associated with linear equations (often called Virasoro or $W$-constraints)
and leads to a $W$-representation of partition function as an action
of an exponentiated differential/difference {\it cut-and-join} operator on unity,
while integrability implies also existence of less trivial non-linear (Hirota-like)
equations, typically quadratic.
Not only are  matrix integrals perfectly suited to the study of this phenomenon,
they allow to further develop it to the concept of {\it superintegrability},
which is a possibility to find a special basis in the space of couplings,
where correlators are simple, deeply factorized quantities.
Despite certain partial results, exact inter-relations inside the quadruple
superintegrability  - Hirota - Virasoro -W-representation
remain an open problem and a subject for future research.

We restrict consideration to Hermitian matrix model in Gaussian background,
where all the four phenomena are well studied,
linear constraints are exactly Virasoro and integrability is of the most familiar
KP/Toda type.
Superintegrability, Virasoro-like constraints and $W$-representation
survive various deformations,
and integrability can be used to describe non-Gaussian backgrounds.
Analogous results can be obtained for other matrices -- unitary, orthogonal, simplectic
and, probably, belonging to arbitrary representations of arbitrary Lie algebras
(including exceptional).
Of special interest are further generalizations to quantum algebras,
currently available  up to DIM, which involve Macdonald polynomials and
their further Kerov and $3$-Schur analogues, which are still under investigation.

\section{Eigenvalue Matrix Models}

The usual starting point in the theory of MAMO is
an integral goes over $N\times N$ Hermitian matrices $H$
\be
Z_G\{t\}:=\frac{1}{V_N}\int dH \exp\left(-\frac{1}{2}\tr H^2 + \sum_{k=0}^\infty t_k\tr H^k\right)
\label{GMint}
\ee
where $V_N$ is the volume of unitary group $U(N)$.
This quantity is considered as a formal series in powers of the couplings $t_k$,
often called "time-variables" -- this terminology comes from the theory of KP hierarchy, see below.
In modern theory (\ref{GMint}) is considered as a particular integral representation of the
{\it non-perturbative partition function} $Z_G\{t\}$, which
depends on the class of fields (one Hermitian matrix), background action (Gaussian),
boundary conditions for fields (zero at infinity) and the set of moduli (coupling constants $t_k$).
In fact all these dependencies are inter-related, what allows to dream about a common non-perturbative
partition function as a kind of a universal object in a universal moduli space,
unifying all field/string theories.

\bigskip

Four approaches are currently available to the study of such quantities.

\bigskip

{\bf 1. Perturbation theory, genus expansion and superintegrability.}
It studies various correlators
\be
\frac{\p^m}{\p t_{i_1}\ldots \p t_{i_m}} Z = \left<\Tr H^{i_1}\ldots \Tr H^{i_m}\right>
\ee
In the case of Gaussian measure $<\ldots>:= \int  dH e^{-\frac{1}{2}\tr H^2} \ldots$
they are easily calculated, satisfy Wick theorem and can be handled by Feynman diagram technique.
Also the topological $1/N$ expansion and t'Hooft calculus in terms of {\it fat} diagrams
is applicable.
Less obvious, there is a special basis in the space of these correlators,
where they acquire an additional peculiar property of {\it superintegrability},
see s.\ref{si} below.

\bigskip

{\bf 2. Ward identities, D-module and Virasoro constraints.}
One can perform an arbitrary change of variables $H\longrightarrow f(H)$ in the  integral,
what leaves it intact
(provided $f(H)$ is polynomial, and the boundary terms at infinity are damped by the
Gaussian factor $e^{-\tr H^2}$).
In the simples basis for the infinitesimal variations $\delta H \sim H^{n+1}$
we obtain
\be
\hat L_n Z_G = \left( -  \frac{\p}{\p t_{n+2}} +
\sum_{k=1}^\infty kt_k \frac{\partial}{\p t_{k+n}} + \sum_{a+b=n}\frac{\p^2}{\p t_a\p t_b}
\right)Z_G = 0,
\ \ \ \ n\geq -1
\ \ \ \ \ \ \frac{\p}{\p t_0} Z_G = NZ_G
\label{virconG}
\ee
what is called the system of  {\it Virasoro constraints}, because $[\hat L_m,\hat L_n] = (n-m)\hat L_{n+m}$
and $\hat L_n$ with $n\geq -1$ form a Borel subalgebra of Virasoro algebra.
This describes $Z_G\{t\}$ as a solution to the compatible system of linear differential equations
with no reference to any integral, i.e. provides a $D$-module description.
The way to restore partition function from Ward identities is also known as
the AMM-CEO topological recursion and is applicable far beyond matrix models.

\bigskip

{\bf 3. $W$-representation.}
The system of Virasoro constraints imply that their special linear combination
\be
\sum_{n=1}^\infty nt_nL_{n-2} Z_G = 0
\ee
is split into a sum of two operators: the dilatation
\be
\hat l := \sum nt_n\frac{\p}{\p t_n}
\ee
and the $2$-descendant of the cut-and-join
\be
\hat W_{2} :=
\sum_{k,l}\left( kl t_kt_l\frac{\p}{\p t_{k+l-2}} + (k+l+2)t_{k+l+2}\frac{\p^2}{\p t_k\p t_l}\right)
\ee
which have a very special commutation relation
\be
\phantom. [\hat l_0, \hat W_{2}] = 2\hat W_{2}
\ee
This implies that
\be
Z_G = e^{\frac{1}{2}\hat W_{2}} \cdot 1
\ee
i.e. represents partition function as a result of action of an operator
on a trivial state.
In other words, non-perturbative partition functions can be substituted
by cut-and-join operators, acting in the moduli space.
Of certain interest is the interplay between this and the $D$-module descriptions.

\bigskip

{\bf 4. Vandermonde, determinant representation and KP-integrability.}
The angular part of the matrix $H$ decouples in (\ref{GMint}) and the integral reduces to that over $N$
eigenvalues $h_a$ of $H$:
\be
Z_G\{t\} = \int \prod_{a<b}^N (h_a-h_b)^2
\prod_{a=1}^N dh_a \exp\left(-\frac{1}{2}h_a^2 + \sum_{k=0}^\infty t_k h_a^k\right)
\ee
Since Vandermonde is actually a determinant,
this provides a determinantal representation for the partition function
\be
Z_G\{t\} = {\rm det}_{a,b} \phi_{a,b}
\label{ZGdet}
\ee
with
\be
\phi_{a,b}:= \int dh \,h^{a+b-2}
\exp\left(-\frac{1}{2}h^2 + \sum_{k=0}^\infty t_k h^k\right)
= \frac{\partial^2}{\partial t_{a-1}\partial t_{b-1}}{\cal H}_G\{t\},
\ee
\be
{\cal H}_G\{t\} = dh
\exp\left(-\frac{1}{2}h^2 + \sum_{k=0}^\infty t_k h^k\right)
\ee
Such determinants with the property
\be
\frac{\p}{\p \phi_c} \phi_{a,b} = \phi_{a+c,b} = \phi_{a,b+c}
\label{phider}
\ee
satisfy bilinear Hirota equations
and are known as KP/Toda $\tau$-functions.
In the particular case of Gaussian Hermitian model there is an additional
symmetry between indices $a$ and $b$, and the relevant one is
the {\it Toda-chain} integrable hierarchy.

The appearance of determinant reflects the relation to the special class
of antisymmetric representations of $\widehat{Sl(N)}$ algebras with unit central charge
(the free-fermion formalism).
Of most interest is generalization of KP/Toda determinantal technique
to arbitrary representations and to other algebras --
bilinear Hirota-like equations are in fact a consequence of
Tanaka-Krein formalism (multiplication of representations)
and are not restricted to fermionic sector.
However, the lack of consensus on notations in general situation
slows down the development of general theory.

\section{Integrability}

The notion of integrability in quantum field theory does not yet have a
commonly accepted definition.
The usual idea in mechanics refers to sufficient number of commuting integrals of motion,
what in general setting leads to the idea of some hidden algebraic structure.
Another. in fact, closely related approach is "a projection method",
which implies that description is adequate
in terms of free fields with possible screening operators (playing the role of projectors).
Still, the universality of integrability for non-perturbative partition functions
is best understood in terms of diffeomorphism invariance of (functional) integrals --
what also refers to a hidden symmetry and group structure, but it is, perhaps,
less straightforward than the naive one, associated with commuting integrals.
Roughly speaking, the naive integrability is formulated in terms of $Sl(N)$ algebras,
while diffeomorphisms are rather related to Virasoro algebra and its further
$W$-generalizations -- and this is what happens at the simplest level of matrix models.
The true structures in true field/string theory are still a mystery --
still there is mounting evidence that they exist in a very broad context
and thus deserve investigation and understanding.

In the present-day theory of matrix models integrability usually appears in a very
restricted traditional context -- of KP/Toda type, associated with free fermions.
Namely, the tau-function is defined as generic Gaussian correlator
\be
\tau\{t|A\} = \left<0\ \Big|\
\overbrace{  e^{\oint t(z)\tilde \psi(z) \psi(z)dz} }^{U\{t\}}\cdot
\underbrace{e^{A_{mn}\tilde\psi_m\psi_n}}_{{\cal O}_A}\ \Big|\ 0\right>
\label{Hireq}
\ee
and the main secret is the "bosonisation property"
\be
\exp\left( \int_{z_1}^{z_2} \tilde\psi(z) \psi(z)dz \right) \sim \tilde\psi(z_1)\psi(z_2)
\label{bosprop}
\ee
It looks especially simple for free fermions and implies a distinguished way to separate
the small set of time-variables $t(z) = \sum t_k z^k$ from the generic "Grassmannian point"
labeled by semi-infinite matrix $A$.
Though elementary, explicit calculations with free fermions are somewhat lengthy and
borrowing, and in the study of matrix models one usually uses some standard corollaries.
Most useful so far were the following five:

\begin{itemize}

\item {\bf Hirota equations.} These are bilinear equations, solved by a $\tau$-function
at arbitrary point $A$:
\be
\int \frac{dz}{z} e^{\sum_k (t_k-t'_k)z^k} \sum_n D^+_n(t)\tau\{t|A\} \cdot D^-_n(t') \tau\{t'|A\}
= 0
\ee
with the shit operators
\be
\sum_n D^\pm_n(t)z^{-n} = \exp\left(\pm\sum_k \frac{1}{kz^k}\frac{\p}{\p t_k}\right)
\ee
Their role is to use (\ref{bosprop}) and generate an operator $\sum_n \tilde \otimes \psi_n$
which commutes with ${\cal O}_A\otimes {\cal O}_A$ and can be moved  to the right in (\ref{Hireq})
to annihilate the vacuum $\sum_n \tilde \psi_n| 0\big> \otimes \psi_n |0\big> = 0$.
It can not be moved to the left, because it does not commute with the bi-evolution operator
$U\{t\}\otimes U\{t'\}$, involving two different sets of time-variables $t$ and $t'$.
As already mentioned, all these arguments can be straightforwardly generalized to arbitrary
representations of arbitrary algebras -- but many technical details were never worked out.

\item {\bf Schur expansion, Plucker relations and Casimir parametrization.}
As a series in powers of $t$, the $\tau$-function can  be expanded over full basis od Schur functions,
labeled by Young diagrams $R$:
\be
\tau\{t|A\} = \sum_R G_R\{A\}\cdot{\rm Schur}_R\{t\}
\ee
then Hirota equations are equivalent to bilinear Plucker relations on the coefficients $K_R$,
Generic solution to plucker relations is parameterized by arbitrary matrix
\be
G_R = \det_{ij} B_{i,R_j-j}
\ee
where $R_j$ is the length if the $j$-th row in the diagram.
This leads to an interesting $\tau$-functions, parameterized by a special basis of Casimir operators
$C_n(R) = \sum_j \Big(\left(R_j-j+\frac{1}{2}\right)^n - \left(-j+\frac{1}{2}\right)^n\Big)$:
\be
\tau\{t,\bar t\big|g\} = \sum_{R} e^{\sum g_n C_n(R)} \cdot {\rm Schur}_R\{t\} \cdot {\rm Schur}_R\{\bar t\}
\label{Castau}
\ee
or, alternatively, by arbitrary function $f$
\be
\tau\{t,\bar t\big|f\} = \sum_R f_R \cdot {\rm Schur}_R\{t\} \cdot {\rm Schur}_R\{\bar t\}
\label{hyptau}
\ee
with $f_R = \prod_{(i,j)\in R} f(i-j)$.
The latter class is sometime called hypergeometric $\tau$-functions.
In the las examples we also added Schur functions of auxiliary times $\bar t$, which complement
KP $\tau$-function to the Toda-lattice one.
They can be eliminated by putting $\bar t_k=\delta_{k,1}$, which, however, leaves behind an essential
$R$-dependent combinatorial factor $d_R:=S_R\{\delta_{k,1}\}$.
For $g_k=0$ in (\ref{Castau}) or $f=1$ in (\ref{hyptau}) Cauchy identity for the Schur functions
provides a trivial $\tau$-function
\be
\tau\{t,\bar t|0\} = \sum_R {\rm Schur}_R\{t\}\cdot {\rm Schur}_R\{\bar t\} =
\exp\left(\sum_k \frac{t_k\bar t_k}{k}\right)
\ee

It is not fully understood what are the implications for integrability
of the substitution of other interesting
systems of polynomials (Jack, Macdonald, 3-Schur, Shiraishi, \ldots)
in place of Schur functions, and what are exact modifications
of free fermion formalism and Hirota equations -- this is one of the interesting questions
in the theory of matrix models, where this kind of deformations is quite straightforward.

\item {\bf Hirota equations in Miwa variables.}
Miwa transform restricts time-variable to an $N$-dimensional subspace
\be
t_k = \frac{1}{k}\,\tr X^k = \frac{1}{k}\sum_{\alpha = 1}^N x_i^k
\ee
This converts Schur functions and their generalizations into orthogonal polynomials
with Vandermonde measures,
\be
 \prod_{\alpha=1}^N \oint_0 \frac{dx_\alpha}{x_\alpha^2}
 \prod_{\alpha<\beta}^N (x_\alpha- x_\beta) \cdot {\rm Schur}_R[X] \cdot {\rm Schur}_{R'}[X^{-1}]
\ee
In this sense matrix a more natural choice for $X$ is unitary or orthogonal rather than symmetric matrix.
Hirota equations can be written in terms of Miwa variables, then they describe a pair of
$\tau$ with one added and one subtracted $x$-variables.
Solutions have peculiar determinant form
\be
\tau[X] = \frac{\det_{\alpha,\beta} \psi_{\alpha}(x_\beta)}{\Delta(X)}
\ee
where Vandermonde in denominator is $\Delta[X] = \prod_{\alpha<\beta}^N (x_\alpha-x_\beta)
= \det_{\alpha,\beta} x_\beta^{\alpha-1}$, and
the functions
\be
\psi_\alpha(x) = x^{\alpha-1}\Big(1+ O(x^{-1})\Big)
\ee
can be arbitrary series with the given asymptotic.
Such representations of KP $\tau$-functions naturally arise in matrix models with
background fields, like Generalized Kontsevich Model.

\item {\bf Forced hierarchies.}
Our original example, the Gaussian model (\ref{ZGdet}) belongs to a special class
of $\tau$ functions, which possess determinant representation with the property
(\ref{phider}).
They are expanded in Schur {\it polynomials} -- a special sub-class of Schur functions,
associated with symmetric representations (single-line Schur diagrams)
and generated by the expansion of $\exp\left(\sum_k t_kz^k\right) = \sum_r S_{[r]}\{t\}z^r$.
In integrability theory they are associated with so called {\it forced} hierarchies
and possess an additional index $N$, which is sometime called the "zero-time" or its conjugate --
a name, natural from the point of view of the equations (\ref{virconG}).

\item {\bf Riemann surfaces.}
An important class of $\tau$-functions appears as $\det \bar\p$ on Riemann surfaces
with singularities or boundaries, where $t_k$ are associated with boundary conditions,
while Grassmannian point $A$ -- with the moduli space.
This is a very important class, which opens a way to study universal moduli space
and non-perturbative string theory, but it is still not used at a large scale in
the theory of matrix models.

\end{itemize}

\section{Superintegrability
\label{si}}

The term {\it superintegrability} usually implies that the problem is fully solvable
in terms of the elementary single-valued functions, without any ramifications
and other transcendentalities.
The typical example are the motions in harmonic and Coulomb potentials.
where the orbits are closed and described by curves of the second order (quadrics).
In these examples there are additional conservation laws, associated with Runge-Lenz vectors,
but most important is the very fact that the motion is exhaustively described
by some well-behaved functions.
In quantum theory superintegrability means that there is a full set of correlators,
which can be explicitly calculated in generic and elementary terms.
In matrix model case the main example are Gaussian models,
where explicitly calculable are averages of Schur functions
${\rm Schur}_R[H] := {\rm Schur}_R\left\{t_k=\frac{1}{k}\tr H^k\right\}$
\be
\left<{\rm Schur}_R\right> :=  \int {\rm Schur}_R[H]\cdot e^{-\frac{1}{2}\tr H^2} dH
= \frac{{\rm Schur}_R[N] \cdot {\rm Schur}_R\left\{\frac{1}{2}\delta_{k,2}\right\}}
{{\rm Schur}_R\{\delta_{k,1}\}}
\ee
Moreover, there is additional set of functions $K_R[H]=S_R[H]+\ldots$
which form an orthogonal set,
\be
\left <K_{R} K_{R'}\right> = \frac{{\rm Schur}_R[N] }{ {\rm Schur}_R\{\delta_{k,1}\}}\cdot\delta_{R,R'}
\ee
and do not change the averages $S_R$ in the following sense:
\be
\left< K_{Q}\ {\rm Schur}_R \right> =
\frac{{\rm Schur}[N]\cdot {\rm Schur}_{R/Q}\left\{\frac{1}{2}\delta_{k,2}\right\}}
{{\rm Schur}_R\{\delta_{k,1}\}}
= \frac{ {\rm Schur}_{R/Q}\left\{\frac{1}{2}\delta_{k,2}\right\}}
{ {\rm Schur}_R\left\{\frac{1}{2}\delta_{k,2}\right\}}
\left<{\rm Schur}_R\right>
\ee

\section{Perspectives}

There are plenty of MAMO examples beyond Gaussian Hermitian model,
where different pieces of above pattern are already revealed and investigated --
what leaves little doubts about their universal nature and applicability.
The main research direction now is the further extension of integrability approach
to new frontiers beyond matrix models, especially to the full fledged field-theory examples.
This already seems to be within reach for non-eigenvalue matrix and tensor models,
for 2d conformal, 3d Chern-Simons, 4d SYM, supersymmetric low-energy gauge
and DIM-controlled brane theories.
The task is to find either quadratic equations for non-pertrurbative partiion functions
or the special basis in the space of correlators, where they are well factorized
and calculable -- what would suit into integrability or superintegrability paradigm respectively
An important technical problem is to extend integrability concept beyond KP/Toda setting
and determinantal (fermionic) examples, what requires a progress in the basics of
representation theory and non-linear algebra.
In the language of quantum field theory these problems are closely related to the subjects like
exact renormalization group, equations for Feynman diagrams and resurgency.

\section*{Further reading}

The literature list below contains a few selected papers on different aspects of integrability story
for matrix models (a solid piece of knowledge) and quantum field theory (projects and dreams).
It contains both old and fresh presentations, which can be compared to understand the
degree and directions of progress in the field.
The list is far from being complete and even representative.
Much more references can be found in the cited papers.

\section*{Acknowledgements}

I appreciate years of common research on these subjects with many colleagues and friends.

This work is partly supported by the grant of the Foundation for the Advancement of Theoretical Physics
“BASIS” and by the joint grants 21-52-52004-MNT-a and 21-51-46010-ST-a.

\end{document}